\documentclass[aps,pra,twocolumn,superscriptaddress]{revtex4-1}
\usepackage{amssymb}
\usepackage{graphicx}
\usepackage{amsmath}
\usepackage{natbib}
\usepackage{siunitx}
\usepackage{textcomp}
\usepackage{bm}

\usepackage{color}
\def\bk{{\bf k}}
\def\bj{{\bf j}}

\def\br{{\bf r}}

\def\la{\langle}
\def\ra{\rangle}
\def\calL{\mathcal{L}}

\def\calU{\mathcal{U}}

\def\e{\epsilon}

\def\pa{\partial}
\def\nn{\nonumber}

\begin{document}

\title{Dynamics and Density Correlations in Matter Wave Jet Emission of a Driven Condensate}
\author{Zhigang Wu}
\affiliation{Shenzhen Institute for Quantum Science and Engineering and Department of Physics, Southern University of Science and Technology, Shenzhen 518055, China}
\affiliation{Center for Quantum Computing, Peng Cheng Laboratory, Shenzhen 518055, China}
\author{Hui Zhai}
\affiliation{Institute for Advanced Study, Tsinghua University, Beijing, 100084, China}

\date{\today }

\begin{abstract} 
Emission of matter wave jets has been recently observed in a Bose-Einstein condensate confined by a cylindrical box potential, induced by a periodically modulated inter-particle interaction (Nature {\bf 551}, 356 (2017)). In this paper we apply the time-dependent Bogoliubov theory to study the quantum dynamics and the correlation effects observed in this highly non-equilibrium phenomenon. Without any fitting parameter, our theoretical calculations on the number of ejected atoms and the angular density correlations are in excellent quantitative agreement with the experimental measurements. The exponential growth in time of the ejected atoms can be understood in terms of a dynamical instability associated with the modulation of the interaction. We interpret the angular density correlation of the jets as the Hanbury-Brown-Twiss effect between the excited quasi-particles with different angular momenta, and our theory explains the puzzling observation of the asymmetric density correlations between the jets with the same and opposite momenta. Our theory can also identify the main factors that control the height and width of the peaks in the density correlation function, which can be directly verified in future experiments. 
\end{abstract}

\maketitle
\section{Introduction}
The ability to manipulate the inter-particle interaction is one of the truly unique aspects of cold atomic systems~\cite{Chin}. In particular, the flexibility to precisely control the interaction in a spatially dependent~\cite{Yamazaki,Yan,Clark} or temporally dependent manner~\cite{Jochim2101,Engels,Pollack,Yukalov,Hulet} leads to novel situations in a quantum many-body system beyond the paradigms of the traditional condensed matter physics. The recently observed matter wave jet emissions from the Chicago group~\cite{Clark:2017aa,Feng} is such an example, where a time-periodic modulation of the interaction strength is carried out in a cold atomic system with an unconventional trapping configuration. 

\begin{figure}[t]
	\centering
	\includegraphics[width=0.43\textwidth]{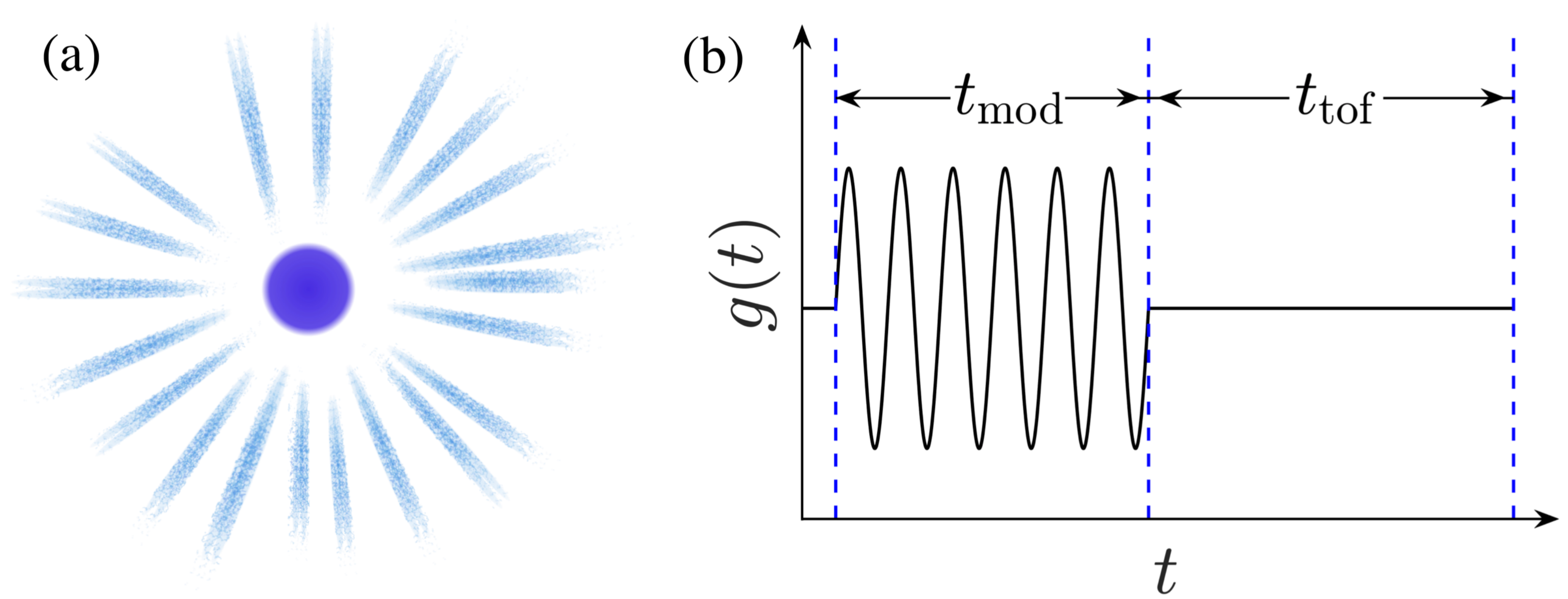}
	\caption{Illustration of the experimental system in Ref.~\cite{Clark:2017aa}. (a) Matter wave jets detected from the a disk-shaped BEC with radius $\rho_0$. (b) Experimental procedure: the interaction is modulated periodically for a time interval $t_\text{mod}$ and the excited atoms travel radially for another time interval $t_\text{tof}$ before detection.  }
		\label{fig:setup}
\end{figure}

In this experimental work~\cite{Clark:2017aa}, a Bose-Einstein condensate (BEC) is confined within a shallow cylindrical box potential and the inter-particle interaction strength is modulated sinusoidally in time, with the modulation frequency much larger than the height of the barrier. Modulation of the interaction naturally leads to excitations from the BEC with energies on the order of the modulation frequency, and as such the shallow barrier cannot prevent the excitations from escaping the trap. After some duration of modulation, bursts of narrow streams of atoms with concentrated density are observed leaving the barrier along the radial direction, as illustrated in Fig.~\ref{fig:setup} (a). Such a phenomenon is termed ``matter wave jet emission" by Ref.~\cite{Clark:2017aa} and, as we shall demonstrate theoretically, provides an ideal platform for the study of many-body correlation effects in a highly non-equilibrium setting. 

The most conspicuous feature of such jet emissions is the fractured density pattern of the ejected atoms in the azimuthal direction in any single measurement. The angular density of the ejected atoms becomes uniform only after averaging over sufficiently many measurements for a given time-of-flight. The underlying azimuthal density pattern in a single image is reflected by the angular density-density correlation, which is found to exhibit peaks at zero and $\pi$ angles even after the average. This is reminiscent of the Hanbury-Brown-Twiss (HBT) effect ~\cite{BROWN:1956aa,Folling:2005aa,Schellekens648}. However, Ref.~\cite{Clark:2017aa} observed a puzzling effect of the correlation, that is the two correlation peaks at zero and $\pi$ angle are highly asymmetric. Since the bosons are initially condensed in the zero-momentum state, if one naturally assumes that the modulation of the scattering length excites pairs of atoms each carrying opposite momentum, one expects that the two correlation peaks to be symmetric~\cite{Clark:2017aa}. This observed asymmetry has even drawn attention from high-energy physics community~\cite{Arratia} where jet emission phenomenon has been studied extensively in high-energy collisions~\cite{Feynman}.

In this paper we apply the time-dependent Bogoliubov theory to study the dynamical behaviour and the density correlations observed in the jet emissions of the BEC. Without requiring any fitting parameter, we find excellent quantitative agreement between our theory and the experimental observations. In fact, among the several recent studies~\cite{Clark:2017aa,Arratia,Fu2018,Chen2018} that attempted theoretical analysis of the experimental results, ours is the only work that achieves such a precise agreement. A key point of our theory is that we obtain the excitations in the angular momentum bases which respect the symmetry of the geometry of the experimental setup. Thus, our theory can attribute the asymmetry of the density correlation function to the destructive interferences between atoms with different angular momenta, and we further predict that such destructive interference processes diminish as the time-of-flight $t_\text{tof}$ increases. Furthermore, based on the distribution of excitations in different angular momenta, we are able to reveal the dependence of the density correlations on the initial condensate size as well as on the driving frequency. All our predictions can be readily verified in future experiments.  

The rest of the paper is organised as follows. In Sec. II we outline the basics of the time-dependent Bogoliubov theory and discuss the applicability of this theory to the problems of interest. In Sec. III, we apply this theory to study the growth of the excitations in a uniform BEC driven by an oscillating magnetic field. We show the existence of a dynamical instability beyond which the number of the excitations grows exponentially. Such an exponential growth underlies the sudden ejections of atoms observed in the jet emission. The phenomenon of the jet emission is fully explored in Sec. IV. We calculate the distribution in angular momentum of the excitations, the total number of excited atoms as well as the angular density-density correlations. The angular density correlation is in particular analysed systematically. All our findings are summarised in Sec. V. 

\section{Time-dependent Bogoliubov theory} 
We consider a trapped and weakly-interacting BEC at zero temperature driven from equilibrium by a time-dependent interaction $g(t)$. The system is described by the time-dependent Hamiltonian ($\hbar = 1$ throughout this paper)
\begin{equation}
\hat H(t) =\int d\br \hat\psi^\dag(\br)\hat h\hat\psi(\br)+ \frac{g(t)}{2}\int d\br \hat\psi^\dag(\br)\hat\psi^\dag(\br)\hat\psi(\br)\hat\psi(\br),
\label{Ha}
\end{equation}
where $\hat h(\br) = -\frac{\nabla^2}{2m} + V_{\rm tr}(\br)$ is the single particle Hamiltonian with $m$ being the atom mass and $V_{\rm tr}(\br)$ the trapping potential. 

Before we consider the time-dependent situation, let's remind ourselves of the time-independent Bogoliubov theory, which describes the collective excitations of the initial condensate at time $t=0$. The initial equilibrium condensate wave function $\Phi_0(\br)$ is determined by the time-independent Gross-Pitaevskii (GP) equation~\cite{Dalfovo} 
\begin{align}
\left [-\frac{\nabla^2}{2m}+V_{\text {tr}}(\br) + g(0)|\Phi_0(\br)|^2\right ]\Phi_0(\br) = \mu \Phi_0(\br),
\label{GP}
\end{align}
where $\mu$ is the initial chemical potential and $\Phi_0(\br)$ normalizes to the total number of atoms $N$. 
In order to describe the collective excitations, we need to obtain the Bogoliubov amplitudes $u_{i,0}(\br)$ and $v_{i,0}(\br)$, which are determined by the Bogoliubov-de-Gennes equations~\cite{Dalfovo} 
\begin{align}
\label{u0}
\calL_0 u_{i,0}(\br) -g(0)\Phi_0(\br)^2 v_{i,0}(\br) &= \e_i u_{i,0}(\br) \\
\calL_0 v_{i,0}(\br) -g(0)\Phi^*_0(\br)^2 u_{i,0}(\br) &= -\e_i v_{i,0}(\br),
\label{v0}
\end{align}
where $\e_i$ is the quasi-particle energy and 
\begin{align}
\calL_0 =\hat h (\br)+2g(0)|\Phi_0(\br)|^2-\mu. 
\end{align}
The so-called fluctuation operator $\delta \hat \psi (\br) \equiv \hat \psi (\br)  - \Phi_0(\br)$ can be expressed in terms of the quasi-particle operators $\hat \beta_j$ and $\hat \beta^\dag_j$
\begin{align}
\label{solheqm1}
\delta \hat \psi(\br)=\sum_j\left [u_{j,0}(\br)\hat \beta_j  -v^*_{j,0}(\br)\hat \beta_j^\dag \right ],
\end{align}
where $\hat \beta_j$ and $\hat \beta^\dag_j$ obey the usual Bose commutation rules. Conversely we have 
\begin{align}
\label{solheqm2}
\hat \beta_j  = \int d\br \left [u_{j,0}(\br) \delta \psi (\br)  + v_{j,0} (\br) \delta \psi^\dag(\br) \right ].
\end{align}

The dynamics of the system for $t> 0 $ can be investigated by means of the time-dependent Bogoliubov theory. In this framework, we consider the time-evolution of the grand-canonical Heisenberg field operator $\hat \psi_K(\br,t) \equiv \hat\calU^\dag(t) \hat \psi(\br)\hat\calU(t)  e^{i\mu t}$, where $\hat \calU(t)$ is the Schr\"{o}dinger evolution operator. We can write
\begin{align}
\hat \psi_K(\br,t)=\Phi_0(\br)e^{i\mu t}+\delta \hat \psi_{K}(\br,t),
\end{align} where  $\delta \hat \psi_{K}(\br,t)$ is the fluctuation operator in the grand-canonical Heisenberg picture. The latter satisfies the following equation of motion
\begin{align}
i \frac{\pa}{\pa t}\delta\hat \psi_{ K}=\calL(\br, t)\delta\hat \psi_{ K}+g(t)\Phi_0^2(\br)\delta \hat \psi^\dag_{ K},
\label{eqpsiK}
\end{align}
where 
\begin{align}
 \calL (\br, t) \equiv \hat h(\br)+2g(t)|\Phi_0(\br)|^2 - \mu.
\end{align}
 The above equation can be solved by the Bogoliubov transformation 
\begin{align}
\delta \hat \psi_K(\br,t)  & = \sum_j\left [u_{j,0}(\br)\hat \beta_{j,K}(t)  -v^*_{j,0}(\br)\hat \beta_{j,K}^\dag(t) \right ] \nn \\
& \equiv \sum_j\left [u_j(\br,t)  \hat \beta_j - v_j^*(\br,t)\hat \beta_j^\dag\right ],
\label{TBT}
\end{align}
where  $ \hat \beta^\dag_{j,K}, \hat \beta_{j,K} $ are the quasi-particle operators in the grand-canonical Heisenberg picture. Substituting the transformation into Eq.~(\ref{eqpsiK}), one finds that the time-dependent Bogliubov amplitudes $u_j(\br,t),v_j(\br,t)$ are determined by the coupled Bogliubov-de Gennes (BdG) equations
\begin{align}
\label{bdgeq1}
i{\pa_t u_j(\br,t)} &=  \calL(\br,t) u_j(\br,t) - g(t) \Phi_0(\br)^2  v_j(\br,t)  \\
i{\pa_t v_j(\br,t)} &= -  \calL(\br,t)  v_j(\br,t) + g(t) \Phi_0^{*}(\br)^2  u_j(\br,t) 
\label{bdgeq2}
\end{align}
where the Bogoliubov amplitudes satisfy the following orthonormal relations
\begin{align}
\int d\br \left [ u_i(\br,t)u^*_j(\br,t) - v_i(\br,t)v^*_j(\br,t) \right ] = \delta_{ij}.
\end{align}
It is clear that the initial conditions for the time-dependent Bogoliubov amplitudes are determined by Eqs.~(\ref{u0})-(\ref{v0}). 

In principle, the above BdG equations are a valid description of the dynamics only when the condensate depletion due to the perturbation is sufficiently small. For large depletions, a set of modified BdG equations together with a generalized time-dependent GP equation are generally needed~\cite{Castin2, Castin1, Gardiner, Billam,Zaremba2009}. Such modifications are particularly necessary for inhomogeneous systems such as harmonically confined condensates~\cite{Gardiner, Billam}. For the systems that we shall consider, however, a simpler approach can be adopted. The first important simplifying factor is that the system of our interest is almost uniform due to the unconventional trapping geometry of the experiment (see also discussion in Sec. IV). In addition, the type of time-dependent perturbation, namely the modulation of the interaction strength, does not break the translational invariance. This is to be contrasted with the more familiar dynamical situation where the perturbation couples to the local density. Finally, we only consider modulation frequencies much higher than the confining potential barrier, such that the excited atoms will immediately depart from the system and no longer interact with the condensate. These considerations suggest that there exists a time-independent condensate mode from which the atoms are continuously depleted during the dynamic process. In other words, we assume that the time-dependent condensate wave function can be approximated by 
\begin{align}
\Phi_0(\br,t) = \sqrt{N_0(t)/N}\Phi_0(\br),
\end{align}
where $N_0(t)$ is the number of the condensed atoms at time $t$. This time-dependent condensate wave function $\Phi_0(\br,t) $ is then used in the BdG equations (\ref{bdgeq1})-(\ref{bdgeq2}) for the calculations of the Bogoliubov amplitudes. The number of condensed atoms $N_0(t)$ can be determined self-consistently using the the conservation of the total number of atoms
\begin{align}
N = N_0(t) +  \int d\br  \left\la \delta \hat \psi^\dag_{K}(\br,t) \delta \hat \psi_{K}(\br,t)\right \ra ,
\label{Nconser}
\end{align}
where  $\la \cdots \ra$ denotes expectation value with respect to the initial ground state. 

Once the condensate wave function and the Bogoliubov amplitudes are determined, all relevant physical quantities can be readily calculated. For example, the number of quasi-particles excited in the $j$-th state at time $t$ is given by
\begin{align}
N_j(t) = \la \hat \beta^\dag_{j,K}(t) \hat \beta_{j,K}(t)  \ra.
\label{nj}
\end{align}
Now using 
\begin{align}
 \hat \beta_{j,K}(t)  = \int d\br \left [u^*_{j,0}(\br)\delta \hat \psi_K(\br,t) + v^*_{j,0}(\br)\delta \hat \psi_K^\dag(\br,t)  \right ]
\end{align}
and Eq.~(\ref{TBT}) in Eq.~(\ref{nj}), we find 
\begin{align}
N_j(t)  = \sum_i \left |  \int d\br \left [v_{j,0}(\br) u_i(\br,t) - u_{j,0}(\br)v_{i}(\br,t) \right ]\right |^2 .
\end{align} 
We see from the above expression that $N_j(t=0) = 0$ as it should be. The total number of excited quasi-particles at $t$ is thus 
\begin{align}
N_{\rm ex}(t) = \sum_j N_j(t). 
\end{align}
We wish to point out here that the total number of excited quasi-particles is generally different from that of atoms not in the condensate. The latter is given by
\begin{align}
N_{\rm nc} (t)& \equiv  \int d\br  \left\la \delta \hat \psi^\dag_{K}(\br,t) \delta \hat \psi_{K}(\br,t)\right \ra  \nn \\ 
& = \sum_j  \int d\br  |v_j(\br,t)|^2.
\end{align}
However, as one can easily check, these two quantities, $N_{\rm ex}(t) $and $N_{\rm nc}(t)$, are identical if the initial condensate is non-interacting, namely if $u_{j,0}(\br)=\varphi_j(\br)$ and $v_{j,0}(\br) = 0$, where $\varphi_j(\br)$ is the eigenstate of $\hat h(\br)$. In fact, they are almost the same for the weakly interacting condensates that we shall consider and for this reason we will not distinguish them in later discussions. 

Similarly one finds that the density and the current density are respectively given by 
\begin{align}
n (\br) &= \la \hat \psi^\dag_K (\br,t)\hat \psi_K (\br,t) \ra  \nn \\
& = n_0(\br,t)  + \sum_j |v_j(\br,t)|^2
\end{align}
and 
\begin{align}
\bj (\br,t) &=  \frac{1}{2m i} \left [\la\hat \psi^\dag_K (\br,t) \nabla \hat \psi_K (\br,t)\ra - c.c.\right ]  \nn \\
& = \frac{1}{2m i} \sum_j \left [ v_j(\br,t) \nabla v^*_{j}(\br,t) - c.c.  \right ].
\end{align}
Finally we consider the equal-time density-density correlation function~\cite{Glauber}
\begin{align}
g^{(2)}(\br,\br';t) \equiv\frac{\left \la \delta \hat n_K (\br,t) \delta \hat n_K (\br',t) \right \ra}{\la \int d\br \delta \hat n_K (\br,t)\ra^2},
\end{align}
where 
\begin{align}
\delta\hat n_K (\br,t) \equiv   \hat \psi^\dag_{K}(\br,t)  \hat \psi_{K}(\br,t)  - n_0(\br, t).
\end{align}  Denoting $\psi_{j-}(\br,t) \equiv \Phi_0(\br,t)\left [ u_j(\br,t) - v_j(\br,t)\right ]$, the correlation function can be written as 
\begin{align}
&g^{(2)}(\br,\br';t)  = \frac{1}{N^2_{\rm nc}(t)}\sum_j \psi_{j-} (\br,t)  \psi^*_{j-} (\br',t) \nn \\
& +\frac{1}{N^2_{\rm nc}(t)} \sum_{jj'} \left \{   |v_j(\br,t)|^2|v_{j'}(\br',t)|^2 \right.  \nn \\
& +\left. \left [ v_j(\br,t)u_{j'}(\br,t)+ v_{j'}(\br,t)u_{j}(\br,t)\right ]u^*_{j}(\br',t)v^*_{j'}(\br',t)\right \}.
\end{align}

\section{A periodically driven uniform condensate}
To gain some insight into how a condensate responds to an oscillating interaction strength, we turn first to the simplest case where the condensate is uniform. The advantage of considering such a case is that its driven dynamics admits some analytical treatment. For a uniform condensate, momentum $\bk$ is a good quantum number and the Bogoliubov amplitudes can be written as $u_\bk(\br,t) = u_\bk(t) e^{i\bk\cdot\br} $ and $v_\bk(\br,t) = v_\bk(t) e^{i\bk\cdot\br} $. Furthermore, a time-dependent interaction strength does not break the translational invariance and thus the condensate wave function remains  spatially uniform, i.e., $\Phi_0(t) = \sqrt{n_0(t)}$, where $n_0$ is the condensate density. With these considerations, Eqs.~(\ref{bdgeq1})-(\ref{bdgeq2}) become
\begin{align}
\label{bdgeqk1}
i{\pa_t} u_\bk(t) &=  \calL(\bk,t)  u_\bk(t) - g(t) n_0 v_\bk(t)  \\
i{\pa_t} v_\bk(t)&= - \calL(\bk,t)  v_\bk(t) + g(t) n_0  u_\bk(t), 
\label{bdgeqk2}
\end{align}
where  $\calL(\bk,t)=\e_\bk+2g(t)n_0-\mu$ with $\e_\bk = \bk^2/2m$ and $\mu = g(0)n_0$. Here $g(t) = 4\pi [a_\text{bg} + a_\text{am}\sin (\Omega t)]/m$ where $a_{\rm bg}$ is the background scattering length and  $a_\text{am}$ is the amplitude of the modulation. Defining $w_\bk(t) = v_\bk(t)/u_\bk(t)$, Eqs.~(\ref{bdgeqk1})-(\ref{bdgeqk2}) can be recast into a single equation
\begin{align}
 i\frac{\pa}{\pa t} w_\bk = - 2\calL(\bk,t) w_\bk + g(t) n_0 \left ( 1+ w_\bk^{2}   \right ),
 \label{wevo}
 \end{align} 
This type of equation, known as Riccati equation, can be easily solved numerically. For some analytical understanding, we consider an initially non-interacting BEC with $a_\text{bg} = 0$ and analyse the growth of the excitations due to the modulation of the interaction. Based on to solutions to Eq.~(\ref{wevo}) we identify the following three stages for the dynamics of the driven BEC: 

(i) Initial Slow Growth: Since initially the number of excitations is very small, we can first neglect the $w^2_\bk$ term in Eq.~(\ref{wevo}). The resulting linear equation can be solved as
 \begin{align}
 w_\bk (t) &= -i\frac{4\pi a_\text{am} n_0}{m} f(t) \int_0^t d t' f^*(t')\sin\Omega t',
 \end{align}
 where $f(t) = e^{2i\e_\bk t - 4i\eta\cos\Omega t}$ with $\eta \equiv {4\pi a_\text{am} n_0}/({m\Omega}) $. For the moment we have neglected the time-dependence of the condensate density. Using the Jacobi-Anger expansion
$ e^{i x \cos\theta} = \sum_{n=-\infty}^\infty i^n J_n(x) e^{in \theta}
$ in the above equation, where $J_n(x)$ is the Bessel function of the first kind, we find that at the resonance energy $\e_\bk = \Omega/2$, $w_\bk (t)$ contains a term that grows linearly in $t$, namely
 \begin{equation}
w_\bk(t) \sim -\frac{2\pi a_{\rm am} n_0}{m}e^{2i\e_\bk t - 4i\eta\cos\Omega t} \left [  J_0 \left (4\eta\right ) + J_2 \left (4\eta\right ) \right ] t. 
\label{wlinear}
 \end{equation}
 This is distinctively different from the case of non-resonance energies where no such growth terms exist and $w_\bk (t)$ remains oscillatory.  Thus at resonance we find at short times 
  \begin{align}
 N_\bk(t) = |v_\bk(t)|^2 = \frac{|w_\bk(t)|^2}{1-|w_\bk(t)|^2} \sim t^2.
 \end{align}
 
(ii) Intermediate Exponential Growth: At resonance energy, $|w_\bk(t)|$ quickly saturates to $|w_\bk| \approx 1$ at the intermediate time scale, and the $w_\bk^2$ term can no longer be ignored.
In this case we can write $w_\bk(t) = |w_\bk(t)|e^{i\varphi_\bk(t)} \approx e^{i\varphi_\bk(t)} $ with $\varphi_\bk(t)$ given by 
 \begin{align}
 \frac{\pa}{\pa t} \varphi_\bk = \Omega + 16 \pi\frac{ a_{\rm am} n_0}{m} \sin(\Omega t) - g_{\rm am} n_0 \sin (\Omega t) \cos \varphi_\bk. 
 \label{phieq}
 \end{align}
This equation suggests that $w_\bk (t) $ is a periodic function with period $T = 2\pi/\Omega$, and without loss of generality, we can expand the imaginary part of $w_\bk (t) $ as $  {\rm Im} w_\bk = -\sum_n\left [ c_n \sin(n\Omega t) + d_n \cos(n\Omega t) \right ]
$. Furthermore, using $\e_\bk = \Omega/2$ and $|w_\bk(t)| \approx 1$,  
we can derive from Eq.~\ref{wevo} the differential equation satisfied by $ N_\bk(t) =|v_\bk|^2$   
\begin{align}
\frac{\pa}{\pa t}  N_\bk(t)
  \approx -8\pi \frac{a_{\rm am} n_0}{m} {{\rm Im} w_\bk}\sin(\Omega t) N_\bk(t).
 \label{diffvk}
 \end{align}
Substituting the Fourier series expansion of ${{\rm Im} w_\bk}$ in the above equation, we can see that $N_\bk(t)$ grows exponentially as
 \begin{align}
 N_\bk(t) \sim e^{4\pi c_1 a_{\rm am} n_0 t/m},
 \label{Nkasym}
 \end{align}
where the coefficient  $c_1$ is positive and generally a function of $\eta$.

(iii) Long Time Saturation: Since the exponential growth rate of the excitations is proportional to the condensate density $n_0$, the growth rate gradually becomes smaller when the condensate is continuously depleted and eventually the number of excitations exhibits a saturation. 

We see from the above analysis that for an initially non-interacting BEC, i.e., $a_{\rm bg} = 0$, the number of excitations experiences an exponential growth for arbitrary modulation amplitude $a_{\rm am}$. We now show that the exponential growth can be attributed to a certain dynamical instability associated with the interaction modulation. We note that analogous instabilities exhibit in BECs for which other parts of the Hamiltonian are modulated periodically~\cite{Dalfovo2005,Creffield2009,Hui2010,Lellouch2017,Lellouch2018,Boulier2018,Nager2018}. Letting $u_\bk(t) = u'_\bk e^{-i \Omega t/2}$ and  $v_\bk(t) = v'_\bk e^{i \Omega t/2}$, and adopting the rotating wave approximation (i.e., terms oscillating at multiples of $\Omega$ are neglected in the equations for $u'_\bk$ and $v'_\bk$), Eqs.~(\ref{bdgeqk1})-(\ref{bdgeqk2}) can be rewritten in the following matrix form
\begin{align}
i \frac{\pa}{\pa t} {\boldsymbol \Lambda}  = M {\boldsymbol \Lambda},
\label{eqchi}
\end{align}
where $\boldsymbol \Lambda \equiv (u'_\bk\quad v'_\bk)^T$, 
\begin{align}
M = \left ( \begin{array}{cc} 
\e_\bk -  \Omega /2 + 4\pi  {a_{\rm bg}n_0} /{m}& -2\pi i {a_{\rm am}n_0}/{m} \\
 -2\pi i { a_{\rm am}n_0}/{m} & -\e_\bk +  \Omega / 2 - 4\pi  {a_{\rm bg}n_0} /{m}
\end{array}  \right )
\end{align}
and we made use of the fact that $\mu = 4\pi  {a_{\rm bg}n_0} /{m}$. Here for simplicity we assume that $n_0$ is constant. The time-dependent behaviour of $u'_\bk$ and $v'_\bk$ are determined by the eigenvalues of the matrix $M$
\begin{align}
E_\bk = \pm \sqrt{(\e_\bk -  \Omega / 2 + 4\pi  {a_{\rm bg}n_0} /{m})^2 - 4\pi^2\left( a_{\rm am}n_0/m\right )^2}.
\end{align}
We see that $E_\bk$ becomes imaginary once the condition 
\begin{align}
|\e_\bk -  \Omega / 2 + 4\pi  {a_{\rm bg}n_0} /{m}| < 2 \pi a_{\rm am}n_0/m
\end{align}
is fulfilled, signalling that the system develops a dynamical instability. It is clear from Eq.~(\ref{eqchi}) that this instability is directly responsible for the exponential growth of the excitations. 
\section{Jet emission}
We now apply the theory outlined in Sec.~II to study the experiment in Ref.~\cite{Clark:2017aa}. There the BEC is trapped by a potential which can be modelled by 
\begin{align}
V_{\rm tr} (\br) = V_{\rm tr}(\rho,\theta,z) = V_h \Theta(\rho-\rho_0) +  m\omega_z^2 z^2/2,
\end{align}
 where $\Theta(x)$ is the Heaviside step function, $\rho_0 $ and $V_h$ are respectively the radius and the height of the cylindrical barrier and $\omega_z$ is the trapping frequency in the vertical direction. At time $t>0$, the atomic scattering length is subjected to a sinusoidal oscillation with the frequency $\Omega \gg V_h$ for a duration of $t_{
\rm mod}$ and is then held at the background value for another duration $t_{\rm tof}$ for the time-of-flight, that is (see Fig.~\ref{fig:setup})
\begin{align}
g(t) = 4\pi [a_\text{bg} + a_\text{am}\sin (\Omega t)\Theta(t_{\rm mod}-t)]/m.
\end{align}
We point out that the main difference between the system considered in this section and that in Sec.~III is that the current system has an open boundary which allows excited atoms to leave the trap. Before solving the BdG equations, we make two observations which will simplify our calculations:

i) Since the background interaction between the atoms is extremely weak with $a_\text{bg} = 5 a_0$ ($a_0$ being the Bohr radius) and the vertical trapping is relatively strong with $\omega_z = 2\pi\times 210 $ Hz, we find that that number of excitations to higher harmonic oscillator states of the vertical trap is negligible during the dynamic process. Thus, in the following calculations we shall consider that all atoms stay in the lowest single-particle eigenstate $\varphi_0(z)$ of the vertical harmonic trap. This is also consistent with the experimental fact that the measured root-mean-square radius of the condensate is in excellent agreement with the harmonic oscillator length along the vertical direction. Such a consideration essentially reduces the system to a two-dimensional one. Furthermore, due to the box trapping potential in the 2D plane, the condensate density is always uniform within the cylindrical barrier and quickly vanishes beyond the radius of the barrier. Thus the time-dependent condensate wave function can be well approximated by
$\Phi_0(\br,t) \approx   \sqrt{n_0(t)} \varphi_0(z)\Theta(\rho_0-\rho)
$, which renders solving the GP equation unnecessary. Here $n_0(t)$ is the density of the condensed atoms per unit area. 

ii) The cylindrical-shaped trap makes it most appropriate to choose the angular momentum eigenstates as the basis to solve the BdG equations. More specifically we consider as the basis the eigenstates $\chi_{ l k}(\br) =   \varphi_{0}(z)  e^{i l \theta}\phi_{l k} (\rho)/\sqrt{2\pi}$ of the single particle Hamiltonian $\hat h (\br) $ with eigenenergy $\e_{lk}$, where $l$ is the angular momentum quantum number, $k$ labels the radial modes of $\hat h (\br)$ and $\phi_{l k} (\rho)$ is the radial wave function.  Due to the angular momentum conservation, we can label the Bogoliubov amplitudes by the index $j \equiv (l,q)$, where $q$ labels the radial modes of the initial equilibrium BdG equations. Thus we can write
\begin{align}
u_{lq} (\br,t) = \sum_{k} \chi_{l k} (\br)U^{(l)}_{kq}(t);\, v_{lq} (\br,t) = \sum_{k} \chi_{l k} (\br)V^{(l)}_{kq}(t). \nn
\end{align}
Using the above expansion, the BdG equations in Eqs.~(\ref{bdgeq1})-(\ref{bdgeq2}) are then converted into two first order differential equations for the matrices $U^{(l)}$ and $V^{(l)}$, which can be solved under the constraint in Eq.~(\ref{Nconser}). 

In the following calculations, all the parameters, including the initial atom number, interaction strengths, trap parameters and the time durations, are chosen to be exactly the same as those used in the experiment.

\subsection{Angular Mode Distribution.} 
\begin{figure}[b]
	\centering
	\includegraphics[width=0.48\textwidth]{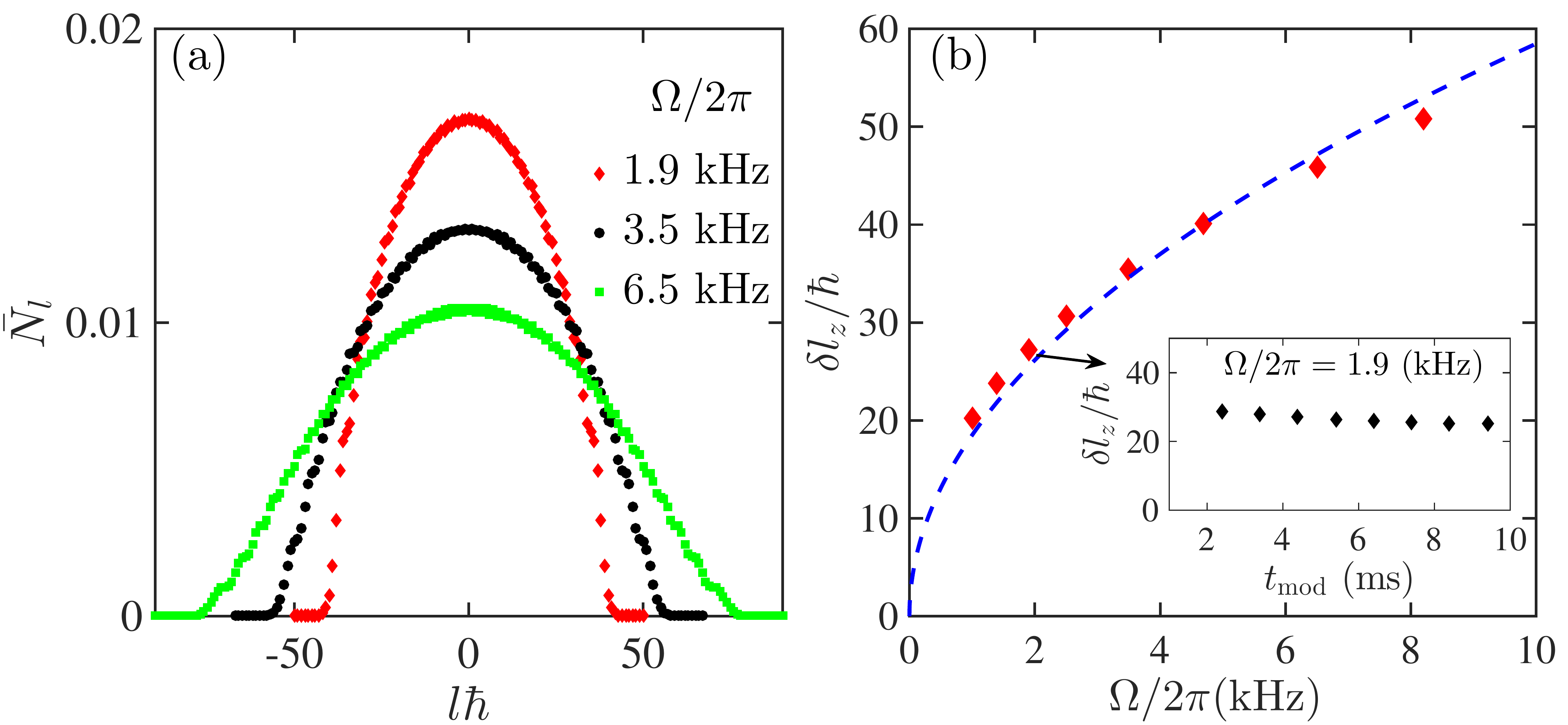}
	\caption{(a) The fraction of atoms $\bar N_l$ with angular momentum $l\hbar$ for three different driving frequencies at $t_{\rm mod} = 4.4$ ms. (b) The angular momentum fluctuation per atom $\delta l_z$ as a function of the driving frequency at $t_{\rm mod} = 4.4$ ms. The inset shows  $\delta l_z$ as a function of $t_{\rm mod}$ for $\Omega = 2\pi\times 1.9$ kHz. The dashed line is a fit by the function $\alpha \rho_0 \sqrt{\Omega}$. Unless specified otherwise, the following parameters are the same for all the calculations: the atom number $N = 30000$, the condensate radius $\rho_0 = 8.5$ {\textmu}m and the modulation amplitude $a_\text{am} = 60a_0$.}
	\label{fig:atomdis}
\end{figure}

First, we consider the mode distribution for the excited atoms. The total number of  atoms excited after a modulation time $t_{\rm mod}$ is
\begin{align}
N_{\rm ex}  = \int d\br \left\la \delta \hat \psi^\dag_{K}(\br,t_{\rm mod}) \delta \hat \psi_{K}(\br,t_{\rm mod})\right \ra = \sum_l N_l,
\label{Nex}
\end{align}
where  
\begin{align}
N_l = \sum_q \int d\br |v_{lq}(\br,t_{\rm mod})|^2 
\end{align}
  is the  number of atoms excited to states with angular momentum $l$. 
 For a sufficiently long modulation time, modes with a range of angular momenta can be significantly occupied. This is shown in Fig.~\ref{fig:atomdis} (a), where the fraction of excited atoms with angular momentum $l$, $\bar N_l \equiv N_l/N_{\rm ex}$, is plotted for several modulation frequencies with an experimental $t_{\rm mod} = 4.4$ ms. For all the occupied radial modes with the same angular momentum, a sharp resonance is found at the energy  $\sim\Omega/2$. Physically,  the resonance at $\Omega/2$ results from the fact that pairs of atoms with opposite angular momentum are excited due to the angular momentum conservation, splitting the total energy quanta $\Omega$ absorbed by the condensate.

Now, the energy of an out-going excited atom can be divided into a rotational part and a radial kinetic part, which together yield  $\Omega/2$ at resonance. The radial kinetic energy decreases as the angular momentum increases, and the occupation of an angular momentum state vanishes when the rotational energy saturates $\Omega/2$. In other words, the rotational energy of an atom with angular momentum $l$ is $\sim l^2/m\rho_0^2 \leq \Omega/2$. Thus the occupation in angular momentum states has a cut-off $l_c$ which determines the width of the distribution and is proportional to $\rho_0\sqrt{\Omega}$.  A more precise characterisation of the width of the angular momentum distribution is  the angular momentum fluctuation per particle 
$
\delta l_z = \sqrt{\sum_l l^2 N_l }/N_{\rm ex}
$,
 which is shown in Fig.~\ref{fig:atomdis} (b) for various driving frequencies employed in the experiments. We expect that $\delta l_z$ is proportional to the cut-off $l_c$ and, as shown in Fig.~\ref{fig:atomdis} (b), it is indeed well described by the function $\alpha \rho_0\sqrt{\Omega}$, where $\alpha$ is a constant.  Furthermore, we find that after some initial transient period $\delta l_z$ depends neither on the modulation time nor on the time-of-flight. As we can see from the inset of Fig.~\ref{fig:atomdis} (b), $\delta l_z$ shows little change for $t_{\rm mod}$ ranging from $2$ ms to $10$ ms. During the time-of-flight, $\delta l_z$ remains unchanged because no more atoms are excited. As we shall see later, this fact determines the robustness of zero correlation peak in the dynamical process of jet emission.

\subsection{Ejected atoms}
\begin{figure}[t]
	\centering
	\includegraphics[width=0.47\textwidth]{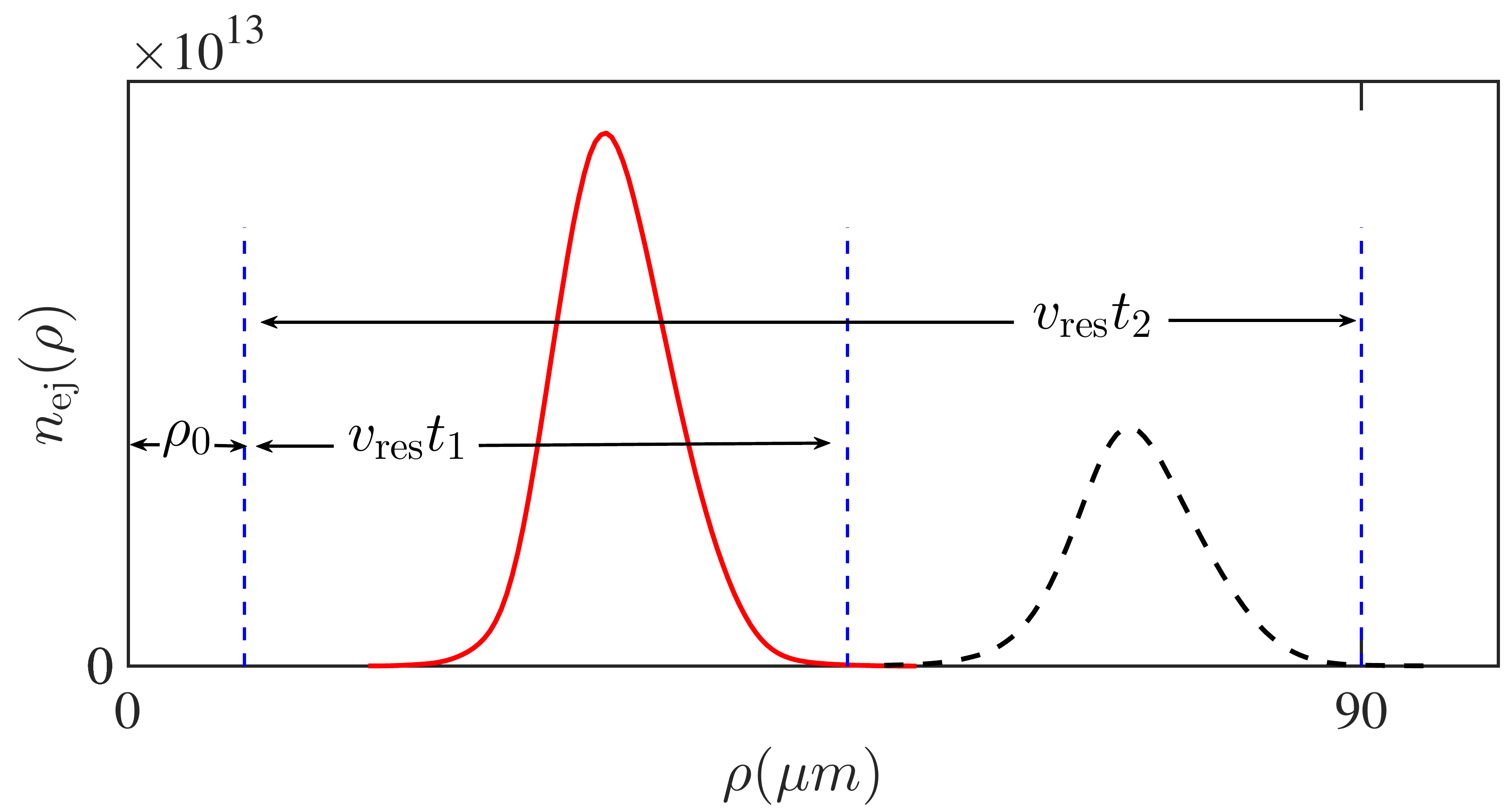}
	\caption{ The radial density profile of the ejected atoms for $t_{\rm mod} = 4.4$ ms at the detection time $t_{1} = 18.4$ ms (red solid) and $t_{2} = 34.2$ (black dashed). For all  calculations, the total atom number is $N = 30000$ and the condensate radius is $\rho_0 = 8.5$ {\textmu}m. Here the modulation amplitude and frequency are  $a_\text{am} = 60a_0$ and $\Omega = 2\pi\times 1900$ Hz respectively.
	}
	\label{fig:density}
\end{figure}

We next calculate the  number of ejected atoms and compare with the experimental measurements.  Since the radial kinetic energy of the excited atoms are mostly on the order of $\Omega/2$, which is much larger than the height of the cylindrical barrier, the excited atoms will escape the barrier after some duration of time-of-flight $t_{\rm tof}$. At $ t = t_\text{mod} + t_{\rm tof}$, the radial profile of the atomic jets is given by 
\begin{align}
n_{\rm ej} (\rho,t) = \sum_{l} |\tilde v_{lq}(\rho,t)|^2
\end{align}
 and shown in Fig.~\ref{fig:atomdis} (b) for $t_{\rm tof} = 14$ ms after a modulation time $t_{\rm mod} = 4.4$ ms. As expected (see Fig.~\ref{fig:density}), the radial distance travelled by the ejected atoms is bounded by $\rho_d = v_{\rm res}\, t$, where $v_{\rm res} = \sqrt{\Omega/m}$ is the resonance velocity for the excited atoms. Although there was no experimental measurement of the radial widths of the jets,  our calculations are consistent with the experimental images of the jet profile. Finally, the total number of ejected atoms at $t$ can be calculated by 
\begin{align}
 N_{\rm ej}(t) = 2\pi \int d\rho \rho \, n_{\rm ej} (\rho,t).
 \label{Nej}
 \end{align}

\begin{figure}[htb]
	\centering
	\includegraphics[width=0.48\textwidth]{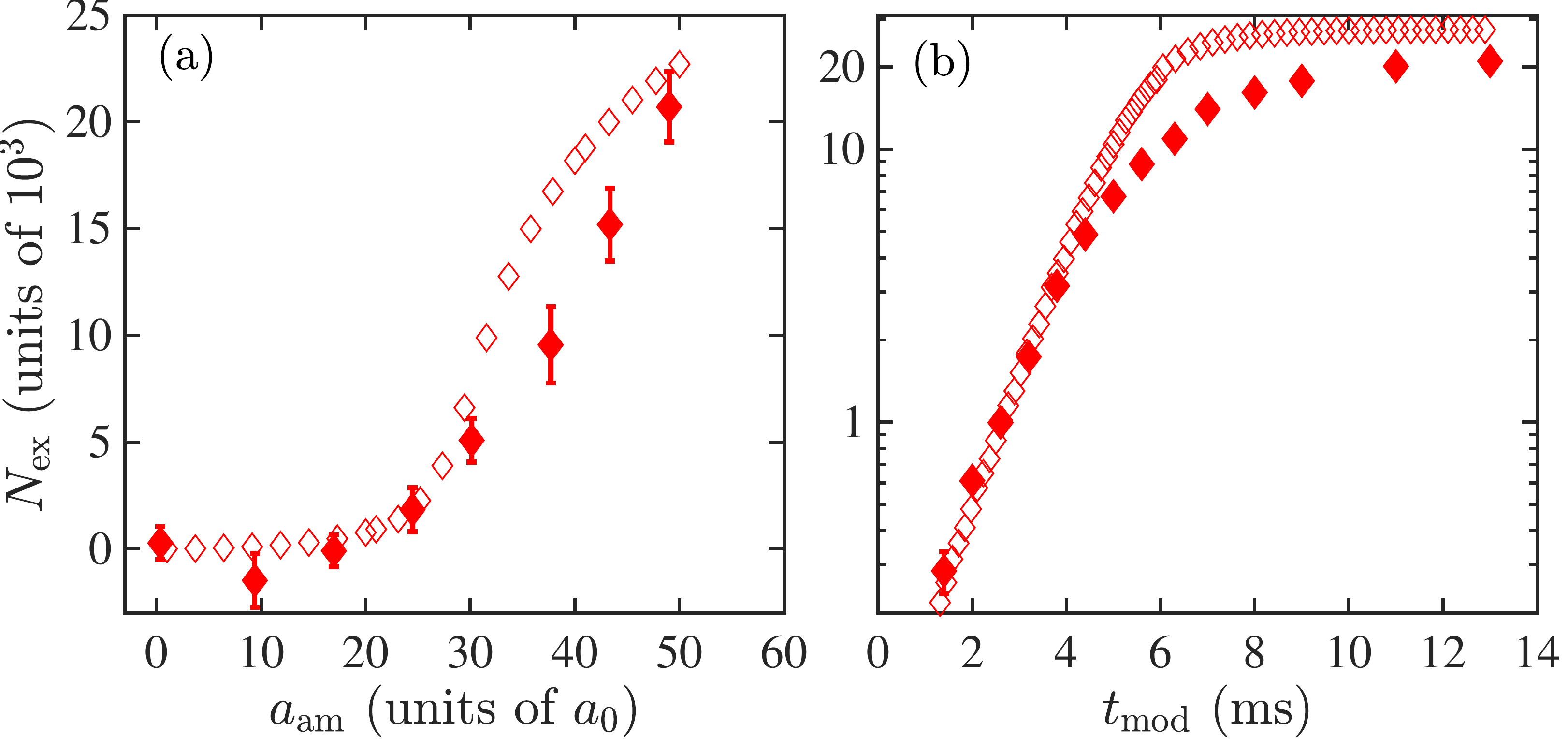}
	\caption{(a):The number of excited atoms as a function of  $a_\text{am}$ for $\Omega = 2\pi\times 4500$ Hz and  $t_{\rm mod} = 25$ ms. (b):The number of excited atoms (logarithmic scale) as a function of  $t_\text{mod}$ for $a_\text{am} = 60 a_0$ and  $\Omega = 2\pi\times 1900$ Hz. The filled symbols are the experimental measurements.}
	\label{fig:cylindrical}
\end{figure}

For a sufficiently long $t_{\rm tof}$, all the excited atoms will leave the barrier after the modulation stops at $t_{\rm mod}$. Thus, the number of ejected atoms $N_{\rm ej}(t)$ detected at time $t=t_\text{mod} + t_{\rm tof}$ simply equals to the number of excited atoms $N_{\rm ex}(t)$ at $t=t_{\rm mod}$. In Fig.~\ref{fig:cylindrical}, we show $N_{\rm ex}(t_{\rm mod})$ calculated from Eq.~(\ref{Nex}) both as a function of $a_\text{am}$ for a fixed $t_{\rm mod}$ and as a function of $t_\text{mod}$ for a fixed $a_{\rm am}$.  We have verified that Eq.~(\ref{Nej}) indeed yields the same results. Without any fitting parameter, the overall agreements between our calculations and  the experimental measurements are again excellent.

\subsection{Angular Density Correlation.} 
\begin{figure}[t]
	\centering
	\includegraphics[width=0.47\textwidth]{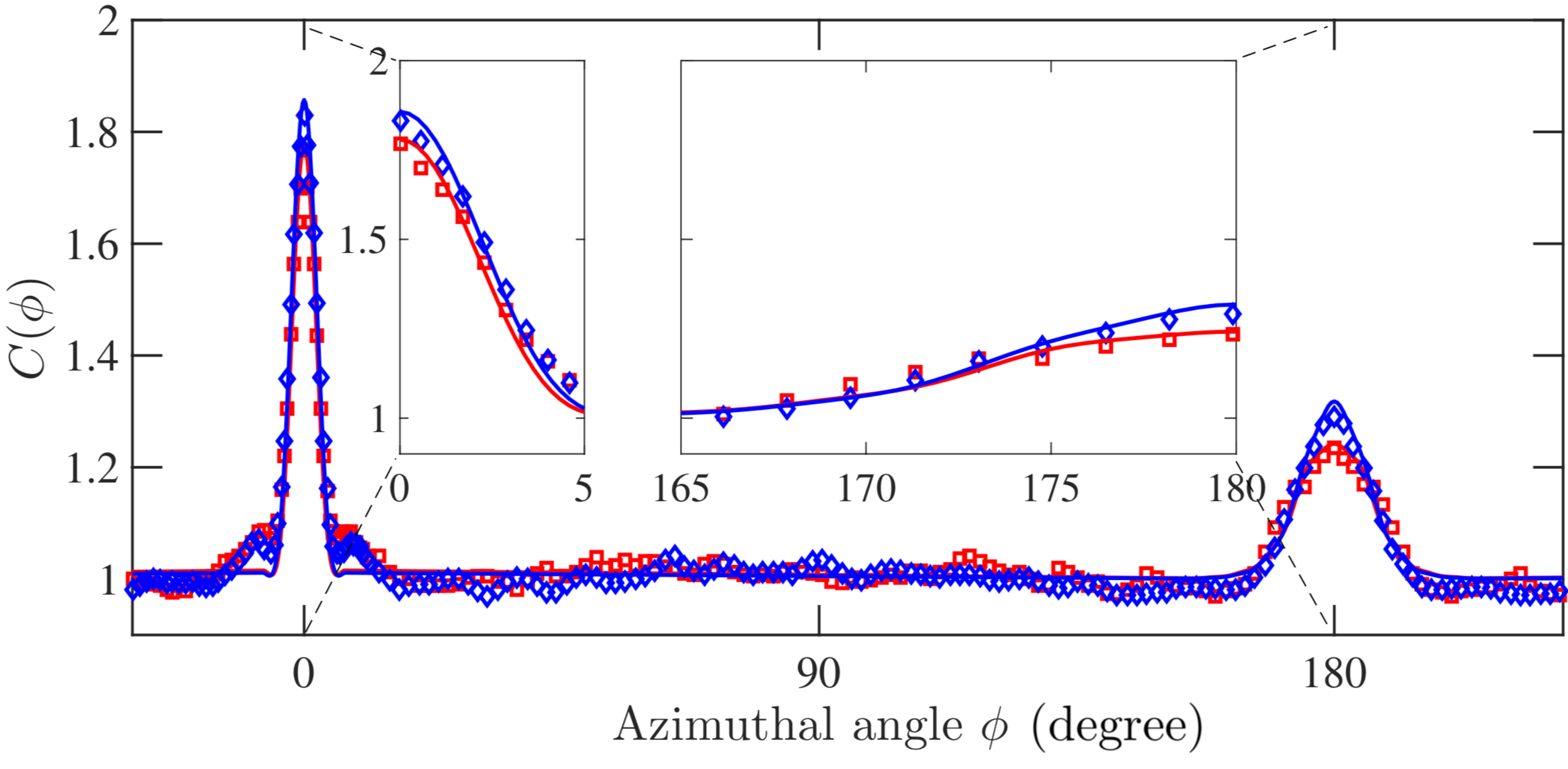}
	\caption{$C(\phi, t)$ calculated with $ t_{\rm mod} = 4.4$ ms (red) and $ 5.6$ ms (blue) for $\Omega = 2\pi\times 1.9$ kHz and $t = 36$ ms. The experimental measurements are all taken at  $t =t_{\rm mod} + t_{\rm tof}  =36$ ms~\cite{footnote} and are shown in symbols in corresponding colors. The insets are expanded views of the correlation peaks.
	}
	\label{fig:correlation1}
\end{figure}

We now consider the angular density correlation function defined in terms of the density correlation function as  
 \begin{align}
 C(\phi,t) =2\pi \int d\theta \int d\rho d z d\rho' dz' g^{(2)}(\br,\br';t),
 \label{corr}
 \end{align}
where in cylindrical coordinates $\br = (\rho,\theta,z)$ and $\br = (\rho',\theta+\phi,z')$. Or more explicitly, the angular correlation function can be written as 
 \begin{align}
 C(\phi,t) = \frac{2\pi\int d\theta\left \la\delta\hat n_K (\theta,t)\delta\hat n_K (\theta+\phi,t) \right \ra}{\left \la \int d\theta \delta\hat n_K (\theta,t)\right \ra^2}, 
 \label{corr1}
 \end{align}
where  $\delta\hat n_K (\theta,t) \equiv  \int dz \int d\rho \rho \, \delta\hat n_K(\br,t)$. Using the Bogoliubov transformation and expressing the Bogoliubov amplitudes as $u_{lq}(\br) = \varphi_{0}(z)  e^{i l \theta} \tilde u_{lq} (\rho) /\sqrt{2\pi}$ and $v_{lq}(\br) =\varphi_{0}(z)  e^{i l \theta} \tilde v_{lq} (\rho) /\sqrt{2\pi}$,  we obtain  
 \begin{align}
 C(\phi,t) = 1 +\sum_{l l'} C_{ll'}(t)e^{i(l+l')\phi}, \label{corr2}
 \end{align}
where 
\begin{align}
C_{ll'}(t) = N_{\rm ex}^{-2} \sum_{q q'} \left [G_{lq,l'q'}+  G_{l'q',lq}\right ] G^*_{lq,l'q'} 
\end{align} 
with $ G_{lq,l'q'}(t) =  \int d\rho \rho \tilde u_{lq}(\rho,t)\tilde v_{l' q'}(\rho,t) $. If we consider situations where the excited atoms have travelled away from the condenstate, the expression in Eq.~(\ref{corr2}) suggests that the experiments are essentially measuring a HBT type correlation between atoms in different angular momentum states. In Fig.~\ref{fig:correlation1}, we compare the correlation function calculated from Eq.~(\ref{corr2}) for two different modulation times to those measured in experiment. The agreement between our theory and the experiment is remarkable, considering that the jet emission is a highly non-equilibrium process. A noteworthy property of the correlation function is the asymmetrical distribution between peaks at $\phi = 0$ and at $\phi=\pi$, which is accurately captured by our theory. This asymmetry can in fact be clearly seen from Eq.~(\ref{corr2}), where all the terms inside the summation contribute constructively for $\phi = 0$ while the terms with odd $l+l^\prime$ contribute destructively to the correlation peak at $\pi$ leading to a reduction of its height. This is in contrast to the theory in Ref. \cite{Clark:2017aa}, which uses the plane wave bases and assumes the conservation of momentum in pair production. Consequently, it always leads to a symmetric distribution between $0$ and $\pi$. In fact, due to the finite size and the disk geometry of the condensate, the momentum conservation is not a good assumption as far as the correlation is concerned.  

\begin{figure*}[ht]
	\centering
	\includegraphics[width=0.97\textwidth]{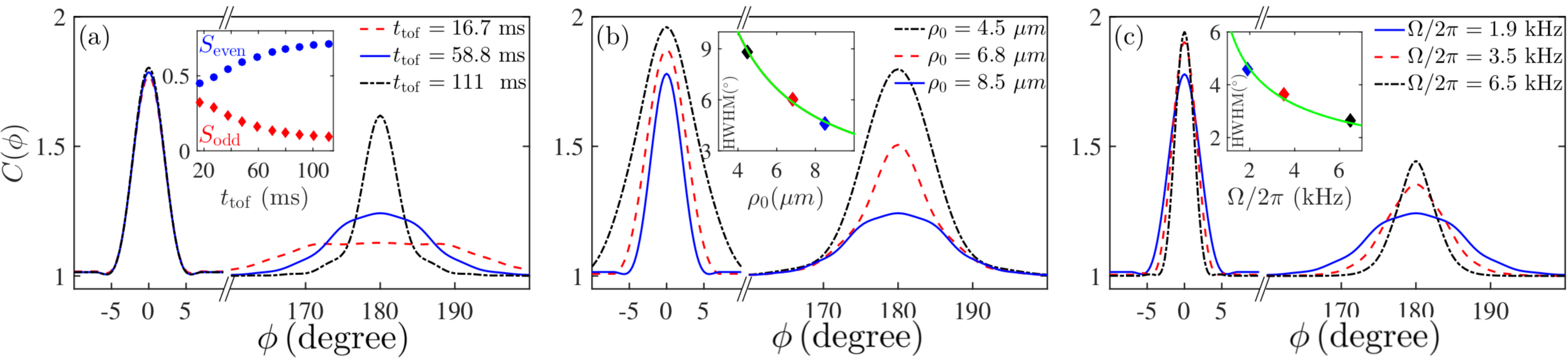}
	\caption{ (a) $C(\phi)$ calculated for different $t_{\rm tof}$ with the same $t_{\rm mod} = 4.4$ ms. The inset shows the behaviour of $S_{\rm even}$ and $S_{odd}$ during time-of-flight (see text). (b) $C(\phi)$ calculated for condensates of different radius (same density as that in (a)) with $t_{\rm mod} = 4.4$ ms and $t_{\rm tof} = 31.6$ ms. The inset shows that the half-width at half-maximum (HWHM) of the zero peak as a function of $\rho_0$ behaves as $\sim 1/\rho_0$ (green line). The driving frequency for both (a) and (b) is $\Omega = 2\pi\times 1.9$ kHz. (c) $C(\phi)$ calculated for different driving frequencies. The inset shows that the HWHM of the zero peak as a function of $\Omega$ behaves as $\sim 1/\sqrt{\Omega}$ (green line). 
	}
	\label{fig:correlation2}
\end{figure*} 
In addition to making comparisons to the experiments, our theory can also reveal systematically how the density correlation depends on the time-of-flight $t_\text{tof}$, the initial condensate radius $\rho_0$ and the driven frequency $\Omega$. For $t_\text{tof}$, as shown in Fig.~\ref{fig:correlation2} (a), the density correlation function is calculated for three increasingly longer $t_{\rm tof}$ with the modulation time fixed at $t_{\rm mod} = 4.4$ ms. We see that while the zero peak remains almost identical during the time-of-flight, the $\pi$ peak begins with a plateau structure at small $t_{\rm tof}$, becomes progressively sharper as $t_{\rm tof}$ increases and eventually becomes similar to the zero peak. 

The robustness of the zero peak during the time-of-flight can be understood from the Heisenberg uncertainty principle  $\delta \phi \,\delta l_z \sim \hbar$. As demonstrated earlier, the angular momentum fluctuation per atom $\delta l_z \sim \alpha\rho_0\sqrt{\Omega}$ depends weakly on $t_{\rm mod}$ or $t_{\rm tof}$, which explains a time-insensitive correlation peak at zero angle. This also leads to the conclusion that the zero peak width is proportional to $1/\rho_0$ and $1/\sqrt{\Omega}$. This is confirmed by numerical calculation presented in Fig.~\ref{fig:correlation2} (b) and (c). In Fig.~\ref{fig:correlation2} (b), we increase the radius of the condensate $\rho_0$ while keeping all other parameters fixed and we find that the width of the zero peak indeed decreases as $1/\rho_0$. Similarly in Fig.~\ref{fig:correlation2} (c), we vary the driving frequency $\Omega$ alone and find that the width of the zero peak decreases as $1/\sqrt{\Omega}$. All these results are consistent with the uncertainty principle. 

We turn next to the time evolution of the $\pi$ peak shown in Fig.~\ref{fig:correlation2} (a). As shown in the inset of Fig. \ref{fig:correlation2} (a), we find that the destructive contribution $S_{\rm odd}= \sum_{l+l'={\rm odd}}C_{ll'}(t)$ gradually decreases toward zero as $t_\text{tof}$ increases and the asymmetry eventually disappears. Meanwhile, $S_{\rm even}= \sum_{l+l'={\rm even}}C_{ll'}(t)$ grows such that the peak at $\phi = 0$ stay almost unchanged. For measurements performed at a sufficiently long $t_{\rm tof}$, the asymmetry disappears because the ejected atoms are in the far field where the condensate can be essentially viewed as a point source.  Thus all the atoms emanating from this point source would have perfect symmetry with respect to the origin of the ejection. In other words, the relative size of the condensate from the perspective of the ejected atoms at the time of detection plays a crucial role in the asymmetry of the correlation function. This is also confirmed by results shown in Fig.~\ref{fig:correlation2} (b) and (c). We see in Fig.~\ref{fig:correlation2} (b) that for the same $t_{\rm tof}$ and driving frequency $\Omega$, the two peaks become increasingly more symmetrical when the condensate radius decreases. In Fig.~\ref{fig:correlation2} (c), we see that the $\pi$ peak also becomes sharper as the driving frequency increases, even though the $t_{\rm tof}$ is held the same. This is because a higher driving frequency translates into a larger escape velocity for the ejected atoms and thus a larger distance from the condensate for the same $t_{\rm tof}$. The situation then is similar to that in Fig.~\ref{fig:correlation2} (a) where the asymmetry in the correlation diminishes as the ejected atoms move further away from the condensate.

\section{Conclusion.} 
In this paper we have studied the quantum dynamics of a BEC for which the interaction strength is modulated periodically in time. We show that such a modulation can result in a dynamical instability, manifesting itself in the exponential growth of the number of the excitations. For a BEC trapped by a cylindrical barrier, such a modulation leads to the phenomenon of jet emission, as first explored experimentally in Ref.~\cite{Clark:2017aa}. In addition to the exponential growth of the ejected atoms observed, intriguing angular density correlation patterns have been found in the experiment. We interpret this angular density correlation observed in the jet emission as the HBT effect of quasi-particles in different angular momenta, excited by a periodical modulation of the interaction of a cylindrically trapped Bose condensate. The average density distribution is uniform along the azimuthal direction since there is no coherence between quasi-particles with different angular momenta. The density-density correlation, however, can exhibit interferences manifested as the HBT effect, and the asymmetry between zero and $\pi$ peaks are due to the difference between the constructive and destructive interferences. Because of the perfect agreement between theory and experiment without any fitting parameter, our theoretical framework can be applied to study future experiments in this and similar settings.    

\textit{Acknowledgement.} This work is supported by China Postdoctoral Science Foundation under Grant No. 2017M620034 (ZW), MOST under Grant No. 2016YFA0301600 (HZ) and NSFC Grant No. 11734010 (HZ).  We thank C. Chin, J. Midtgaard, L. Feng, L. Clark for very helpful discussions. We also want to thank the Chicago group for providing the measurement data presented in Ref.~\cite{Clark:2017aa}. 

\bibliographystyle{apsrev4-1}

\end{document}